\documentclass[12pt, a4paper]{article}
\pdfoutput=1

\usepackage{subcaption}
\usepackage[T1]{fontenc}
\usepackage{float}
\usepackage{latexsym}
\usepackage{amsmath,amstext,amssymb,amsfonts,
amscd,bm,array,multirow,amsbsy,mathrsfs,amsthm,calc}
\usepackage{t1enc}
\usepackage{indentfirst}
\usepackage{pb-diagram}
\usepackage{graphicx}
\usepackage{enumerate}
\usepackage{color}
\usepackage[all]{xy}
\usepackage{hyperref}
\hypersetup{colorlinks=false,pdfborderstyle={/S/U/W 0}}
\usepackage[usenames,dvipsnames]{xcolor}
\usepackage{url}
\usepackage{upgreek}

\theoremstyle{plain}

\theoremstyle{definition}

\theoremstyle{remark}

\newcommand{\be}{\begin{equation*}}
\newcommand{\ee}{\end{equation*}}
\newcommand{\ben}{\begin{equation}}
\newcommand{\een}{\end{equation}}
\newcommand{\beqa}{\begin{eqnarray*}}
\newcommand{\eeqa}{\end{eqnarray*}}
\newcommand{\beqan}{\begin{eqnarray}}
\newcommand{\eeqan}{\end{eqnarray}}
\newcommand{\nn}{\nonumber}

\def\i{\mathbf{i}}

\def\C{\mathbb{C}}
\def\R{\mathbb{R}}

\def\H{\mathbb{H}}
\def\D{\mathbb{D}}

\def\dd{\mathrm{d}}

\def\cC{\mathcal{C}}

\def\cG{\mathcal{G}}
\def\cH{\mathcal{H}}

\def\cM{\mathcal{M}}

\def\cX{\mathcal{X}}

\newcommand{\A}{{\Bbb{A}}}
\def\mD{\mathbb{D}}
\def\mA{\mathbb{A}}

\newcommand{\eqdef}{\stackrel{{\rm def.}}{=}}

\def\grad{\mathrm{grad}}

\def\tvarphi{\widetilde{\varphi}}

\def\Re{\mathrm{Re}}
\def\Im{\mathrm{Im}}

\def\tV{\widetilde{V}}
\def\hV{\widehat{V}}

\def\tv{\widetilde{v}}
\def\CP{\mathbb{C}\mathbb{P}}

\begin{document}

\begin{center}
\hfill \\

{\Large{Cosmological flows on hyperbolic surfaces\footnote{Acknowledgements: 
The work of E.M.B. was mainly supported by the Romanian Ministry of Research and Innovation, 
grant PN 18090101/2018, and partly from the ICTP--SEENET-MTP project NT-03 ``Cosmology - 
Classical and Quantum Challenges'' for her participation to BW2018. 
The work of C. I. L. was supported by grant IBS-R003-D1.}}
\\ }

\vskip 1cm

{\bf Elena Mirela Babalic${}^a$\footnote{mbabalic@theory.nipne.ro} and Calin Iuliu Lazaroiu${}^{b\,a}$\footnote{calin@ibs.re.kr}\\
\vskip 0.5cm  {\it ${}^a$ Horia Hulubei National Institute for Physics and Nuclear Engineering (IFIN-HH), Bucharest-Magurele, Romania\\
${}^b$ Center for Geometry and Physics, Institute for Basic Science, Pohang 37673, Republic of
Korea}
}

\end{center}

\vspace{0.3cm}

\noindent {\bf Abstract}
{\small We outline the geometric formulation of cosmological flows for FLRW
models with scalar matter as well as certain aspects which arise in
their study with methods originating from the geometric theory of
dynamical systems. We briefly summarize certain results of numerical
analysis which we carried out when the scalar manifold of the model is
a hyperbolic surface of finite or infinite area.}

\vspace{0.5cm}






\section{Introduction}
 \vglue-10pt
 \indent
 
The standard cosmological model of matter coupled to gravity involves
an FLRW space-time with flat spatial section, with classical
dynamics of matter fields reduced to spatially homogeneous
configurations. The evolution of such models with respect to
cosmological time is governed by generally covariant systems of ODEs
which can be studied \cite{WE,Coley} using the theory of dynamical
systems, which in this context is most powerful when formulated
geometrically \cite{Palis}.

In most such applications up to date, the matter content of the
cosmological model was taken to be relatively simple and 
was assumed not to carry any interesting `internal' geometry. In
particular, scalar inflationary models studied in the dynamical systems framework have largely been assumed to have kinetic terms described by scalar
manifolds which are simply-connected and flat.  However,
arguments from string theory and supergravity suggest that the matter
content present in the universe immediately after the Big Bang could
be quite complicated and in particular that it may involve multiple
scalar fields whose kinetic terms are described by rather general
scalar manifolds, which need not be flat or simply connected. In this
context, there is currently substantial interest in so-called
multifield $\alpha$-attractor models \cite{K1,K2}, whose most general
incarnation involves an arbitrary hyperbolic scalar manifold
\cite{genalpha}. When applied to such models, the dynamical systems
approach leads to highly non-trivial problems which have deep
connections with several branches of mathematics, while posing
interesting challenges regarding the statistical interpretation of
cosmological dynamics. These aspects are spectacularly illustrated
already by the case of scalar two-field models whose scalar manifold
is a hyperbolic surface. Indeed, the cosmological dynamics of such
models relates not only to uniformization theory, Fuchsian groups and
number theory (as already pointed out\footnote{See \cite{S1,S2,S3,S4}
  for a different but related approach.} in \cite{genalpha}) but also,
at least in certain limits, to the theory of hyperbolic and
Morse-Smale flows as well as to certain aspects of ergodic theory and
of non-equilibrium classical statistical mechanics. In references
\cite{elem} and \cite{modular}, we illustrated the complexity of
cosmological dynamics for models whose scalar manifolds are certain
hyperbolic surfaces of finite and infinite area and discussed some of
the first predictions made by such models for cosmological observables.

\section{Cosmological flows}
\vglue-10pt \indent

The cosmological equations of an FLRW model can be reduced to
equations governing the time evolution of spatially
homogeneous field configurations by eliminating the conformal scale
factor $a(t)$ using the Friedmann constraint. The resulting
differential equations (which involve only matter
variables\footnote{The matter variables describe spatially homogeneous
  configurations of matter fields in the FLRW universe and hence
  depend only on the cosmological time.}) define a dynamical system on
the relevant space of states, whose flow we call the {\em cosmological
  flow} of the model. For models involving only scalar fields, these
differential equations form a geometric system of coupled second order
ODEs governing the evolution of scalar variables, which is described
by a point moving inside the scalar manifold $\cM$. In such models,
the cosmological flow is defined on the tangent bundle $T\cM$, which
is the space of states of the associated dynamical system.

\subsection{Scalar triples and cosmological equations of motion} 

\noindent As explained in \cite{genalpha}, cosmological scalar field
models of FLRW type with flat and simply-connected spatial section are
parameterized by the choice of a {\em scalar triple}
$(\cM,\cG,V)$. The {\em scalar manifold} $(\cM,\cG)$ is a generally
non-compact but complete Riemannian manifold which determines the
kinetic terms of the scalar fields, while the {\em scalar potential}
$V$ is a smooth and everywhere non-negative real-valued function
defined on $\cM$. After eliminating the conformal scale factor $a(t)$
using the Friedmann constraint, the evolution of the model is
described by smooth curves $\varphi:I\rightarrow \cM$ (with $I\subset
\R$ an interval) which satisfy the following geometric second order
autonomous ODE, where $t$ is the cosmological time:
\begin{equation}
\label{eom}
\nabla_t \dot{\varphi}(t)+ \left[||\dot{\varphi}(t)||_{\varphi(t)}^2+2V(\varphi(t))\right]^{1/2}\dot{\varphi}(t)+ (\grad V)(\varphi(t))=0~~.
\end{equation}
Here the norm, covariant derivative and gradient are taken with
respect to the scalar manifold metric $\cG$. The second term arising
in the left hand side of \eqref{eom} (known as the {\em Hubble
friction term}) involves the {\em Hubble function} $H_\varphi$ of the
curve, which is defined through:
\begin{equation}
H_\varphi(t)\eqdef \cH(\varphi(t),\dot{\varphi}(t))~~,
\end{equation}
where the {\em absolute Hubble function} $\cH$ of the scalar triple
$(\cM,\cG,V)$ is defined on $T\cM$ through the formula:
\begin{equation}
\cH(u)\eqdef \frac{1}{3}\left[||u||_{\pi(u)}^2+2V(\pi(u))\right]^{1/2}=\frac{1}{3}\sqrt{2E(u)}~~(u\in T\cM)~~.
\end{equation}
Here $\pi:T\cM\rightarrow \cM$ is the bundle projection and
$E=\frac{9}{2}\cH^2\in \cC^\infty(\cM,\R)$ is the {\em Hubble energy}:
\begin{equation}
E(u)\eqdef \frac{1}{2}||u||_{\pi(u)}^2+V(\pi(u))~~.
\end{equation}
The Hubble energy and absolute Hubble function are continuous on $T\cM$ and
smooth on the slit tangent bundle $\dot{T}\cM\eqdef T\cM\setminus 0$
(where $0$ denotes the image of the zero section of $T\cM$).
The following one-parameter deformation of \eqref{eom}:
\begin{equation}
\label{eomdef}
\epsilon \nabla_t \dot{\varphi}(t)+ \left[||\dot{\varphi}(t)||_{\varphi(t)}^2+2V(\varphi(t))\right]^{1/2}\dot{\varphi}(t)+ (\grad V)(\varphi(t))=0
\end{equation}
interpolates between the geodesic equation $\nabla_t
\dot{\varphi}(t)=0$ of $(\cM,\cG)$ (which is recovered for
$\epsilon \rightarrow \infty$) and the equation
$\left[||\dot{\varphi}(t)||_{\varphi(t)}^2+2V(\varphi(t))\right]^{1/2}\dot{\varphi}(t)+
(\grad V)(\varphi(t))=0$ (which is obtained for $\epsilon\rightarrow 0$).  As
explained in \cite{genalpha}, the latter is equivalent with
the gradient flow equation $\frac{\dd \varphi(q)}{\dd q}+ (\grad
V)(\varphi(q))=0$ of the scalar triple $(\cM,\cG,V)$ through the
change of parameter $t\rightarrow q$ defined through:
\begin{equation}
\dd t=3H_\varphi(q)\dd q~~.
\end{equation}
Hence the cosmological equation of motion (which corresponds to
$\epsilon=1$) `sits between' the geodesic and gradient flow equations.
Unlike the geodesic equation (which is the Euler-Lagrange equation of
the free particle Lagrangian $L(u)\eqdef
\frac{1}{2}||u||^2_{\pi(u)}$), the cosmological equation of motion
need {\em not} admit a Lagrangian description (see \cite{Carroll} for
a discussion of this point in the simple case of one-field models).
Using the interpolation provided by \eqref{eomdef}, one can develop
two perturbation expansions for the dynamics of the model, namely the
{\em gradient flow expansion} (whose leading approximation was already
discussed in \cite{genalpha}) and the {\em geodesic flow expansion}. 
Besides these two, there are numerous other perturbation expansions 
which can be considered for such models. 

\subsection{The cosmological semispray and cosmological flow}
\vglue-10pt \indent

\noindent Since the second order ODE \eqref{eom} is geometric, it is
equivalent with the flow equation:
\begin{equation}
\label{dyn}
\dot{\gamma}(t)=S(\gamma(t))~~(\mathrm{where}~~\gamma:I\rightarrow T\cM)
\end{equation}
of a vector field $S$ defined on $T\cM$ which satisfies 
the {\em semispray condition} (see \cite{Bucataru}):
\begin{equation}
\pi_\ast(S_u)=u~~,~~\forall u\in T\cM
\end{equation}
and which we call the {\em cosmological semispray} of the scalar
triple $(\cM,\cG,V)$.  More precisely, a curve $\gamma:I\rightarrow
T\cM$ is a solution of \eqref{dyn} iff it coincides with the complete
lift\footnote{The {\em complete lift} of a curve $\varphi:I\rightarrow
  \cM$ is the curve $\tvarphi:I\rightarrow T\cM$ defined through
  $\tvarphi(t)\eqdef (\varphi(t),\dot{\varphi}(t))\in T\cM$ for any
  $t\in I$.} $\tvarphi$ of a solution $\varphi:I\rightarrow
\cM$ of \eqref{eom}. One can show that the following relation holds:
\begin{equation}
\label{S}
S=S_0+3\cH C+(\grad V)^v~~,
\end{equation}
where $S_0$ is the geodesic spray of $(\cM,\cG)$, $C$ is the Liouville
vector field of $T\cM$ and $X^v\in \cX(T\cM)$ denotes the vertical
lift of a vector field $X\in \cX(\cM)$. Hence the classical dynamics
of the model is described by the {\em cosmological dynamical system}
defined by the vector field \eqref{S} on $T\cM$, whose flow we call
the {\em cosmological flow} of the scalar triple $(\cM,\cG,V)$. The
state space $T\cM$ of this dynamical system is the space of positions
and velocities of the spatially-homogeneous scalar field distributions
described by the curve $\varphi$. When $\cH$ and $\grad V$ are small,
the cosmological semispray \eqref{S} can be viewed as a vertical
perturbation of the geodesic spray of $(\cM,\cG)$.

The cosmological flow of a general scalar triple can be extremely
complicated and its proper study requires deep ideas from the theory
of dynamical systems, asymptotic analysis and singular perturbation
theory. One useful perspective on this flow is provided by scale
analysis, a.k.a. by `renormalization group' \footnote{In our context,
scale analysis is applied to a classical dynamical system, unlike
Wilson's famous application of such methods to quantum field
theories.}  techniques. In the models discussed here, one can show
that the IR limit corresponds to a reparameterized gradient flow,
while the UV limit corresponds to the geodesic flow. In particular,
the regime of most interest for inflation (which is the `slow motion',
i.e. the IR regime) can be understood by studying the gradient flow
and gradient flow expansion. The UV and IR regimes of the model are
markedly different, as illustrated quite dramatically by the case
$\dim\cM=2$.

\section{Cosmological flows on hyperbolic surfaces} \vglue-10pt
\indent
 
As mentioned above, the cosmological flow of a general scalar triple
$(\cM,\cG,V)$ can be extremely complicated, so it is useful to
consider the case of two-dimensional scalar manifolds. In this
situation, $\cM$ is a (generally non-compact and not simply connected)
smooth surface which we shall denote by $\Sigma$.

The special case when $\cG=3\alpha G$ with $\alpha$ a positive
parameter and $G$ a complete hyperbolic metric on $\Sigma$ of
Gaussian curvature $-1$ produces a {\em generalized two-field
$\alpha$-attractor model} in the sense of reference
\cite{genalpha}. Such models are particularly interesting since, under
mild assumptions on the scalar potential, they have universal behavior
for certain special cosmological trajectories which are close to the
Freudenthal ends of $\cM$.  More precisely, it was shown in
\cite{genalpha} that for scalar potentials having `good' asymptotic
behavior at the ends and in the slow-roll approximation for certain
cosmological trajectories $\varphi$ located close to the ends, the
naive one-field truncation produces the same values of the spectral
index ${\bf n_s}$ and tensor to scalar ratio ${\bf r}$ as ordinary
one-field $\alpha$-attractors, thus being in good agreement
with current observations:
\begin{equation} 
{\bf n_s}\approx 1-\frac{2}{{\cal N}},~~~{\bf r}\approx \frac{12\alpha}{{\cal N}^2}~~,
\end{equation} 
where ${\cal N}\eqdef \int_{t_i}^{t_f}H_\varphi(t)\dd t$ is the number
of efolds. In fact, much stronger universality arguments can be made
for such models using dynamical systems techniques.

Models based on hyperbolic surfaces are also interesting from a
mathematical perspective, given their deep connection to Fuchsian
groups and number theory (which stems from Poincar\'e's uniformization
theorem) and the special behavior of their geodesic flow. For example,
it is well-known that the geodesic flow of a hyperbolic surface of
finite area is ergodic and mixing.

In \cite{elem} and \cite{modular}, we performed a numerical study of
cosmological flows for certain non-compact hyperbolic
surfaces. Reference \cite{elem} considered cosmological flows on
elementary hyperbolic surfaces (namely the Poincar\'e disk $\mD$, the
hyperbolic punctured disk $\mD^\ast$ and the hyperbolic annuli
$\mA(R)$), while \cite{modular} studied the case of the hyperbolic
triply-punctured sphere $Y(2)$.

As explained in \cite{genalpha}, the equations of motion of any
two-field generalized $\alpha$-attractor model can be lifted from
$\Sigma$ to the Poincar\'e half-plane $\H$ by using the covering map
$\pi_\H:\H\rightarrow \Sigma$ which uniformizes $(\Sigma,G)$ to $\H$.
This allows one to determine the cosmological trajectories
$\varphi(t)$ by projecting to $\Sigma$ the trajectories $\tvarphi(t)$
of a ``lifted'' model defined on $\H$, which is governed by the
following system of second order non-linear ODEs:
\begin{eqnarray}
\label{elplane}
&& \ddot{x}-\frac{2}{y}\dot{x}\dot{y} +\frac{1}{M_0} \left[3\alpha \frac{\dot{x}^2
+\dot{y}^2}{y^2}+2\tV(x,y)\right]^{1/2}\!\!\dot{x}+\frac{1}{3\alpha} y^2 \partial_x\tV(x,y)=0 \\
&& \ddot{y}+\frac{1}{y}(\dot{x}^2-\dot{y}^2)+\frac{1}{M_0} \left[3\alpha \frac{\dot{x}^2+\dot{y}^2}{y^2}
+2\tV(x,y)\right]^{1/2}\!\!\dot{y}+\frac{1}{3\alpha} y^2\partial_y\tV(x,y)=0~~.\nn
\end{eqnarray}
Here $M_0=\sqrt{\frac{3}{2}} M$ (where $M$ is the reduced Planck
mass), while $x=\Re\tau$, $y=\Im \tau$ are the Cartesian coordinates
on the Poincar\'e half plane with complex coordinate $\tau$ and
$\tV\eqdef V\circ \pi_\H:\H\rightarrow \R$ is the {\em lifted
  potential}. Let $u=\pi_\H(\tau)$ be the complex coordinate on
$\Sigma$. Let $u_0$ be any point of $\Sigma$ and let $\tau_0\in \H$ be
chosen such that $\pi_\H(\tau_0)=u_0$. An initial velocity vector
$v_0=\dot{u}_0 \in T_{u_0}\Sigma$ and its unique lift
$\tv_0=\dot{\tau}_0\in T_{\tau_0}\H$ through the differential of
$\pi_\H$ at $\tau_0$ are related through:
\begin{equation}
v_0=(\dd_{\tau_0}\pi_\H)(\tv_0)~~.
\end{equation}
Writing $\tau=x+\i y$ and $\tau_0=x_0+\i y_0$, we have
$\tv_0=\tv_{0x}+\i \tv_{0y}$. A cosmological trajectory on
$\Sigma$ with initial condition $(u_0,\tau_0)$ can be written as
$\varphi(t)=\pi_\H(\tvarphi(t))$, where $\tvarphi(t)=x(t)+\i y(t)$ is
the
solution of the lifted system \eqref{elplane} with initial conditions:
\begin{equation}
x(0)=x_0~~,~~y(0)=y_0~~\mathrm{and}~~\dot{x}(0)=\tv_{0x}~~,~~\dot{y}(0)=\tv_{0y}~~.
\end{equation}

In the next subsections, we limit ourselves to presenting examples of 
trajectories on $\mD^\ast$ and $\A(R)$ found in \cite{elem} for the
corresponding globally well-behaved scalar potentials which lift to
the following smooth function:
\begin{equation}
\label{hatV}
\hV_+(\psi)=M_0\cos^2\frac{\psi}{2}~
\end{equation}
 written in spherical coordinates $(\psi, \theta)$ on $S^2$.
The trajectories were obtained by numerical computation of solutions
\eqref{elplane} on the Poincar\'e half-plane, followed by projection
to $\Sigma$ through the explicitly-known uniformization maps. 

\subsection{Cosmological trajectories on the hyperbolic punctured disk $\D^\ast$}
\label{subsec:TrajPuncDisk}
 \vglue-10pt
 \indent

\begin{figure}[H]
\centering
\begin{minipage}{.47\textwidth}
\centering
\includegraphics[width=0.92\linewidth]{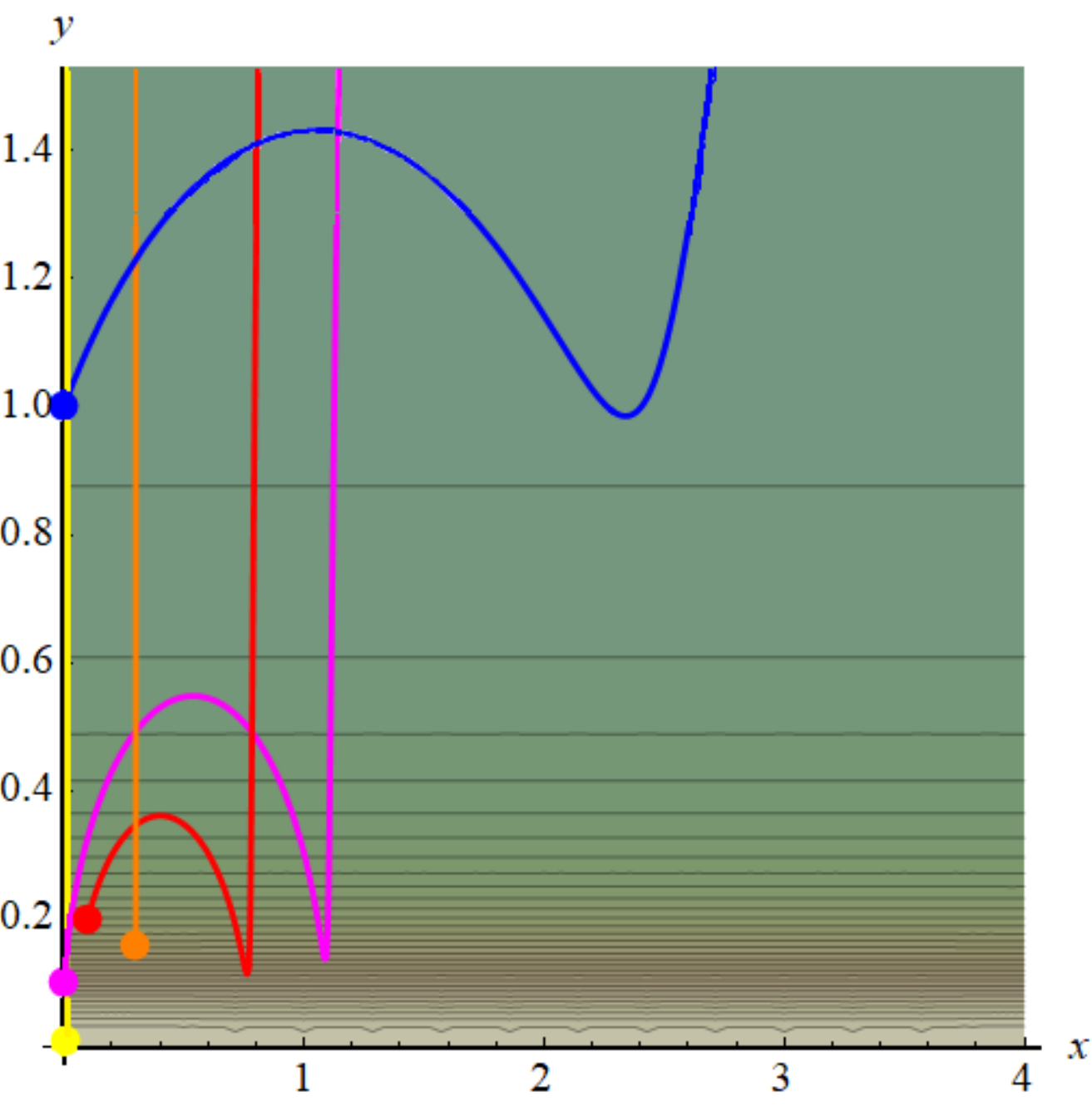}
\subcaption{Trajectories for $\tV=\tV_+$ on $\H$.}
\label{fig:PuncDiskPhiPlus}
\end{minipage}\hfill
\begin{minipage}{.47\textwidth}
\centering
\includegraphics[width=0.91\linewidth]{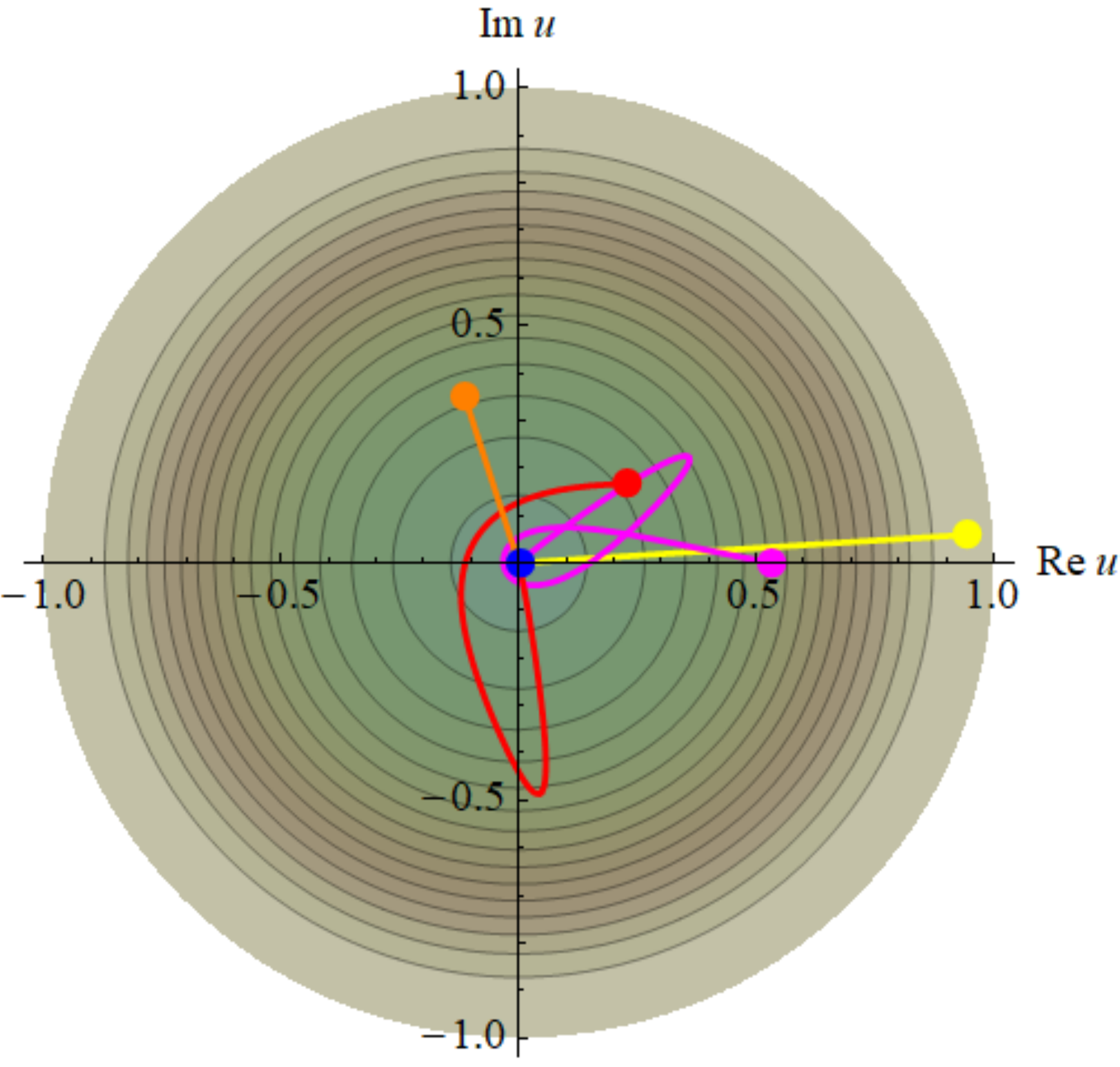}
\vskip 2mm
\subcaption{Projection on $\mD^\ast$ of the trajectories shown at the left.}
\label{fig:PuncDiskPhiPlusProj}
\end{minipage}
\caption{Numerical solutions for $V=V_+$ and $\alpha=\frac{M_0}{3}$.}
\label{fig:PuncDiskPhiPlusAll}
\end{figure}

\noindent Figure \ref{fig:PuncDiskPhiPlusAll} shows five lifted
trajectories on the Poincar\'e half-plane $\H$ and their projections
to $\mD^\ast=\{u\in\C~|~0<|u|<1 \}$ (which is endowed with its unique
complete hyperbolic metric) for the globally well behaved scalar
potential:
\begin{equation}
 V_+=M_0\frac{1}{1+(\log|u|)^2}~~,
\end{equation}
with the initial conditions listed in Table \ref{table:InCond1}.
Since $V_+$ has a minimum at the cusp end (which corresponds to the
center of the disk), it produces an attractive force toward the cusp,
which acts as counterbalance to the repulsive effect of the hyperbolic
metric. Out of the five trajectories, the one showed in yellow produces 55 efolds,
which fits the observationally favored range of 50-60 efolds.

\noindent In all figures, we show in the background the level sets of the potential
$\tV=\tV_+$ on $\H$, respectively $V=V_+$ on $\Sigma$, where dark
green indicates minima of the potential and light brown indicates
maxima.

\begin{table}[H]
\centering
\caption{Initial conditions for five trajectories on the Poincar\'e half-plane.}
\label{table:InCond1}
\begin{tabular}{|c|c|c|}
\hline
trajectory & $\tau_0$ & $\tilde{v}_0$ \\
\hline
\hline
 orange  & $0.3+0.159\,\i$  & $0$ \\
\hline
yellow & $0.01+0.009\,\i$  & $0$\\
\hline
 red  & $0.1+0.2\,\i$ & $2+3\,\i$  \\
\hline
blue & $\i$ & $1+\,\i$  \\
\hline
 magenta & $0.1\,\i$  & $1.3+7\,\i$\\
\hline
\end{tabular}
\end{table}

\subsection{Cosmological trajectories on the annuli $\A(R)$}
\label{subsec:annulus}
 \vglue-10pt
 \indent

\begin{figure}[H]
\centering

\begin{minipage}{.47\textwidth}
\centering 
\includegraphics[width=.76\linewidth]{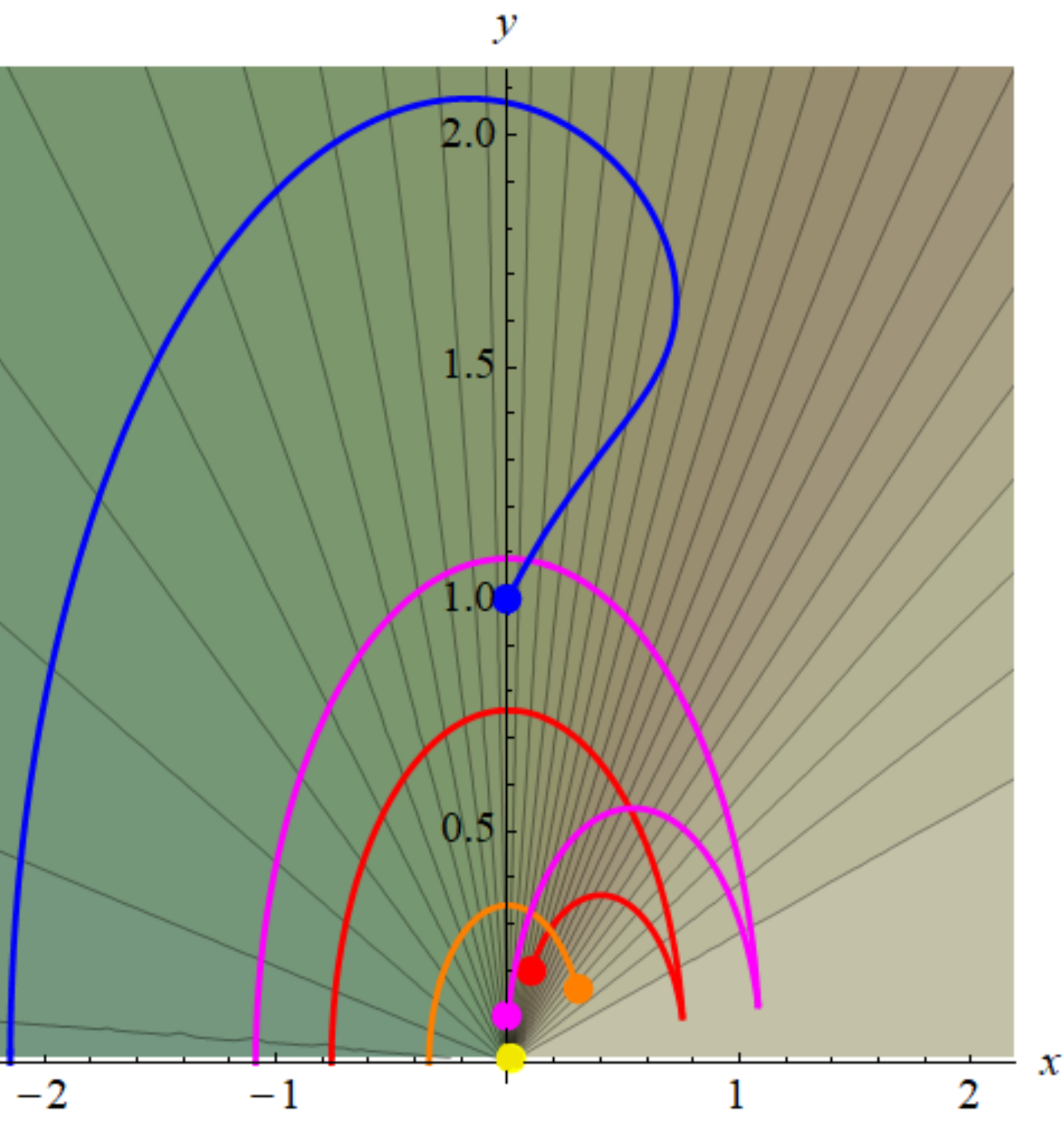}
\subcaption{Trajectories on $\H$ for $\tV=\tV_+$. }
\label{fig:AnnulusPhiPlus}
\end{minipage}
\hfill
\begin{minipage}{.47\textwidth}
\centering
\includegraphics[width=.76\linewidth]{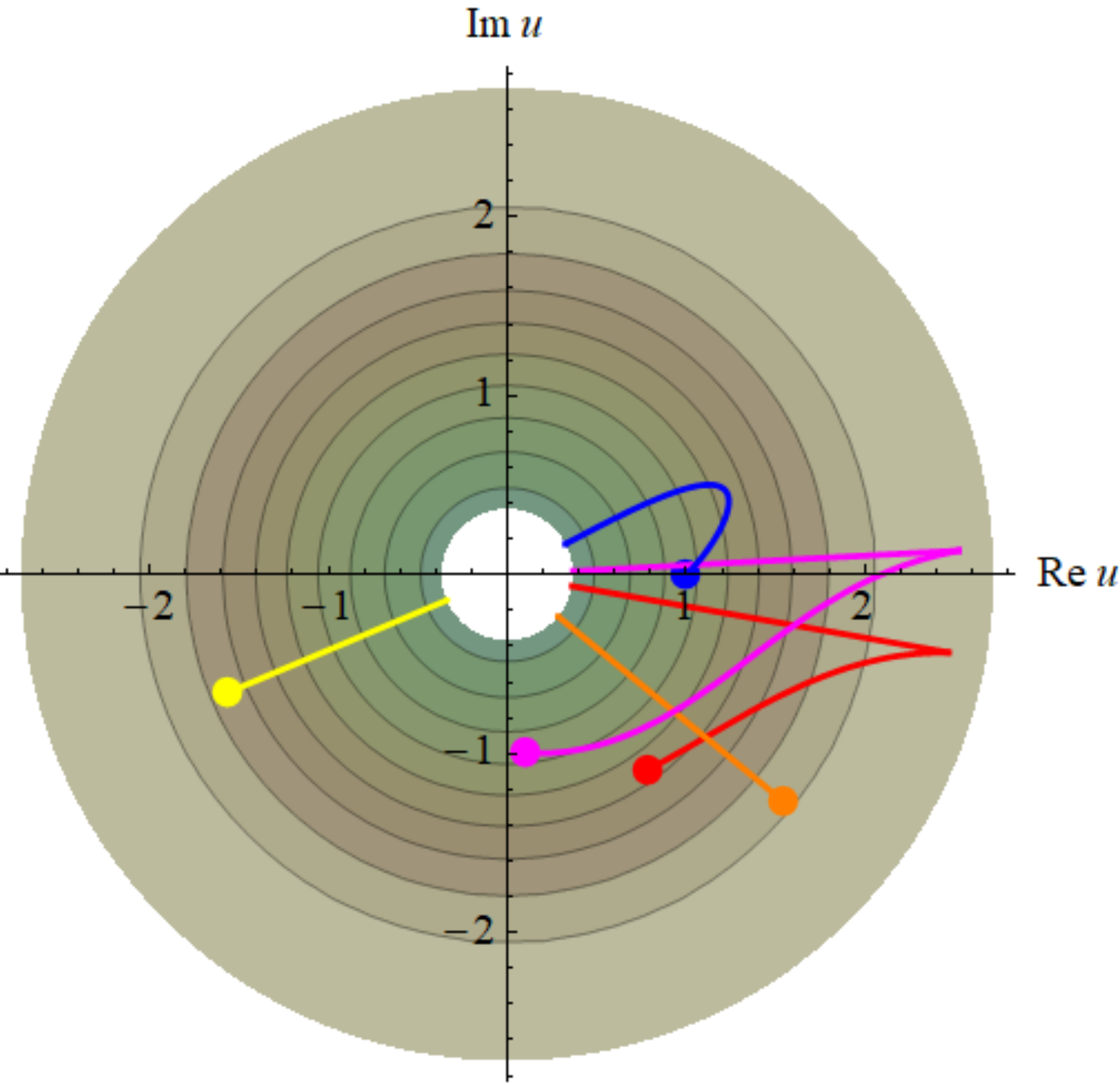}
\subcaption{Projection to $\A(R)$ of the trajectories shown at the left.}
\label{fig:AnnulusPhiPlusProj}
\end{minipage}

\caption{Five trajectories for $V=V_+$, $\alpha=\frac{M_0}{3}$ and 
$R=e$, with initial conditions of Table \ref{table:InCond1}.}
\label{fig:AnnulusPhiPlusAll}
\end{figure}

Figure \ref{fig:AnnulusPhiPlusAll} shows five trajectories (orange,
yellow, red, blue and magenta) on the Poincar\'e half-plane $\H$ and
on the annulus $\A(R)\eqdef\{u\in\C ~|~\frac{1}{R}<|u|<R \}$ (which
is endowed with its complete hyperbolic metric of modulus $\mu=2\log
R$) for $\alpha=\frac{M_0}{3}$, $R=e$ and scalar potential:
\begin{equation}
V=V_+=M_0\frac{1}{1+\Big[\log\frac{R-\frac{1}{R}}{|u|-\frac{1}{R}} \Big]^2}~~,
\end{equation}
with the same initial conditions of Table \ref{table:InCond1}. 
Here the potential induces an attractive
force toward the inner funnel end, making some trajectories to turn at some point in $\A(R)$ 
and evolve back toward the inner funnel end.

\noindent In this case the orange and yellow trajectories give 76 and
74 efolds respectively, while the remaining three trajectories do not
start in the inflationary region of the chosen potential.

\section{Conclusions}
\vglue-10pt \indent

Cosmological models with multiple scalar fields described by general
scalar triples $(\cM,\cG,V)$ have not been studied systematically from
a global perspective. Such models lead to geometric dynamical
systems defined by a certain semispray on the tangent bundle $T\cM$,
whose flow `interpolates' in an appropriate sense between the geodesic
flow of $(\cM,\cG)$ and the gradient flow of $(\cM,\cG,V)$. This flow
becomes particularly interesting when the complete metric $\cG$ has
negative sectional curvature. In particular, it relates to 
deep aspects of asymptotic analysis and ergodic theory. 

When the scalar manifold is a surface $\Sigma$ endowed with a metric
of the form $\cG=3\alpha G$ with $\alpha$ a positive parameter and $G$
a complete hyperbolic metric defined on $\Sigma$, the associated
cosmological model is a two-field generalized $\alpha$-attractor model
in the sense of \cite{genalpha}. The cosmological flow of such models
can already be very intricate, especially when $(\Sigma,G)$ has finite
hyperbolic area. Numerical studies as well as arguments based on the
gradient flow approximation indicate \cite{elem,modular,unif} that such
models can be compatible with current observational constraints.

The epistemological falsifiability of scalar cosmological models is
limited by the largely arbitrary choice of the scalar potential, a
problem which is only compounded in multi-field models.  As such, it
is natural to look for criteria which could constrain the choice of
$V$. A natural way to achieve this is to require that the model admits
a non-generic symmetry. In reference \cite{Noether}, we studied
two-field cosmological $\alpha$-attractors with $(\Sigma,G)$ an
elementary hyperbolic surface, determining those scalar potentials for
which such models admit a `separated' Noether symmetry. This approach
can be extended to more general hyperbolic surfaces and could serve as
one avenue for further constraining such models.

\end{document}